\documentclass[11pt,twoside]{article}
\usepackage{./asp2014}

\aspSuppressVolSlug
\resetcounters

\begin{document}

\title{SALT spectroscopy of evolved massive stars}
\author{A.Y.~Kniazev$^{1,2,3}$, V.V.~Gvaramadze$^{3,4}$ and L.N.~Berdnikov$^{3,4,5}$ 
\affil{$^1$South African Astronomical Observatory, PO Box 9, 7935, South Africa; \email{akniazev@saao.ac.za}}
\affil{$^2$Southern African Large Telescope Foundation, PO Box 9, 7935 Observatory, Cape Town, South Africa;}
\affil{$^3$Sternberg Astronomical Institute, Lomonosov Moscow State University, Moscow 119992, Russia;}
\affil{$^4$Space Research Institute, Russian Academy of Sciences, Profsoyuznaya 84/32, 117997 Moscow, Russia; \email{vgvaram@mx.iki.rssi.ru}}
\affil{$^5$Astronomy and Astrophysics Research Division, Entoto Observatory and Research Center, PO Box 8412, Addis Ababa, Ethiopia;
\email{berdnik@sai.msu.ru}}}

\paperauthor{A.Y. Kniazev}{akniazev@saao.ac.za}{}{South African Astronomical Observatory}{SALT}{Cape Town}{Western Cape}{7935}{South Africa}
\paperauthor{V.V.~Gvaramadze}{vgvaram@mx.iki.rssi.ru}{}{Lomonosov Moscow State University}{Sternberg Astronomical Institute}{Moscow}{}{119992}
{Russia}
\paperauthor{L.N.~Berdnikov}{berdnik@sai.msu.ru}{}{Lomonosov Moscow State University}{Sternberg Astronomical Institute}{Moscow}{}{119992}
{Russia}

\begin{abstract}
Long-slit spectroscopy with the Southern African Large Telescope (SALT) of central stars of mid-infrared nebulae 
detected with the {\it Spitzer Space Telescope} and {\it Wide-field Infrared Survey Explorer} ({\it WISE}) led 
to the discovery of numerous candidate luminous blue variables (cLBVs) and other rare evolved massive stars. With 
the recent advent of the SALT fibre-fed high-resolution \'echelle spectrograph (HRS), a new perspective for the study 
of these interesting objects is appeared. Using the HRS we obtained spectra of a dozen newly identified massive stars.
Some results on the recently identified cLBV Hen\,3$-$729 are presented. 
\end{abstract}

\section{Introduction}

The high angular resolution of the modern infrared (IR) telescopes, the {\it Spitzer Space Telescope} and 
{\it Wide-field Infrared Survey Explorer} ({\it WISE}), allowed us to reveal numerous candidate evolved 
massive stars through the detection of their mid-IR circumstellar nebulae \cite[e.g.][]{Gv10, Gv12}. 
Follow-up optical and IR spectroscopy of these stars showed that among them the most widespread are the candidate 
and bona fide luminous blue variables (LBVs). A census of newly identified stars of this type is given in table\,1 
of \cite{Gv17}. About five dozens of stars with mid-IR nebulae were observed in 2010--2015 with the 
Southern African Large Telescope (SALT), using the Robert Stobie Spectrograph \cite[RSS;][]{Burgh2003} in the 
long-slit mode. In most cases, the obtained spectra cover the spectral range of 4200$-$7300~\AA. The spectral 
classification of these stars is given in table\,1 of \cite{KG2015}. In 2015 the high-resolution \'echelle 
spectrograph (HRS) for the SALT become available, which opened new possibilities 
for more detailed study of the newly identified massive stars with mid-IR nebulae.
With the HRS, we obtained spectra of a dozen such stars. Below, we describe the procedure of reduction of 
the HRS spectra and present some results on the recently identified cLBV Hen\,3$-$729.

\section{Reduction of the HRS spectra}
\label{obs}

HRS is a dual beam, fibre-fed \'echelle spectrograph \citep{Ba08, Br10, Br12, Cr14}.
It could be used in the low (R=14\,000--15\,000), medium (R=40\,000--43\,000) and high (R=67\,000--74\,000)
resolution modes (hereafter, LR, MR and HR modes). Up to now, we carried out twenty \'echelle observations in 
the LR and MR modes of a dozen stars with mid-IR nebulae. 

Primary reduction of the HRS data, including overscan correction, bias subtractions and gain correction, was done 
with the SALT science pipeline \citep{Cr10}. Spectroscopic reduction of the HRS data was carried out using
our own HRS pipeline created with use of the standard \mbox{MIDAS} contexts \texttt{feros} \citep{St99} and 
\texttt{echelle} \citep{Ba92}. The current version of this pipeline automatically reduces data for both 
blue and red arms for all HRS modes. It includes the next standard steps for all modes: (1) positions for 36 spectral 
orders for the blue arm data and 33 orders for the red arm data were found using spectral flats frames; (2) the 2D 
background was determined and subtracted from all frames; (3) the straightened \'echelle spectrum was extracted for
both fibres from all types of frames (flats, arcs and object) using the standard mode with cosmic masking and the 
optimum extraction algorithms; (4) for the object and sky fibres the blaze function was removed from the science frame 
through division to extracted spectrum of spectral flat; (5) the procedure found $\sim$1000 emission lines in the 
extracted arc spectrum of which $\sim$450 lines were finally automatically identified with requested level of tolerance 
to build a 2D dispersion curve with the final mean rms of 0.003, 0.005 and 0.007~\AA\ for the HR, MR and LR modes,
respectively. This step was done independently for the object and sky fibres; (6) all extracted orders were rebinned 
into linear wavelength steps; (7) the wavelength calibrated sky fibre orders were subtracted from the object fibre 
orders; (8) all orders for the object fibre were merged into a 1D file. The spectral ranges of 3900--5500~\AA\ 
for the blue arm and $5400-8900$~\AA\ for the red arm were covered with a final reciprocal dispersion of 
$\sim0.04$~\AA\ pixel$^{-1}$. First results of our HRS observations were presented in a paper on the new 
Galactic bona fide LBV MN48 \citep{KGB2016}.

\section{Some results}
\label{results}

In 2015--2016, we obtained the HRS spectra of a dozen stars with mid-IR nebulae. For six stars (including the 
newly identified bona fide LBVs Wray\,16-137, WS1, MN44 and MN48) two or more spectra were taken. The obtained data 
will be used to derive the fundamental parameters of the observed stars, and, potentially, to measure their radial and 
rotational velocities. Comparison of the HRS spectra of WS1 with the RSS ones \cite[presented in][]{Kn15}
showed that this LBV has entered in the hot state in 2015, which is manifested in re-appearance of strong emission 
lines of He\,{\sc i} in the spectrum of this star. 

One of the stars observed with the HRS is the cLBV Hen\,3$-$729 (also known as ALS\,2533). It was classified 
as a possible Wolf-Rayet star by \citet{St71}. Using the {\it WISE} survey, we discovered a circular shell around this star, 
while the SALT RSS spectroscopy (carried out in 2013--2015) showed that Hen\,3$-$729 has a rich emission spectrum
typical of hot LBVs. The {\it WISE} 22\,$\mu$m image of the nebula around Hen\,3$-$729 and one of the RSS spectra of this star were 
presented for the first time in \citet{KG2015}. 

We classify Hen\,3$-$729 as a cLBV because this star does not show major photometric and spectral variability -- the defining characteristics
of the bona fide LBVs \citep{Hu94}. This is illustrated in Fig.\,1, which plots the light curve of Hen\,3$-$729 in the $V$ band in 1999--2016. 
The filled (red) dots are the photometry from the All Sky Automated Survey \cite[ASAS;][]{Po97}, while the open (blue) stars correspond to the 
CCD photometry obtained with the 76\,cm and 1\,m telescopes of the South African Astronomical Observatory during our observing runs 
in 2009--2016. Fig.\,1 shows that the brightness of Hen\,3$-$729 experiences a quasi-periodical variability with an amplitude not exceeding 
few tenths of magnitude. Similarly, comparison of the RSS and HRS spectra of Hen\,3$-$729 (see Fig.\,2) does not reveal significant changes 
in their appearance on a time scale of several years. A detailed study of Hen\,3$-$729 and other stars observed with the HRS
is underway and will be presented elsewhere.

\articlefigure[angle=-90,width=0.9\textwidth,clip=]{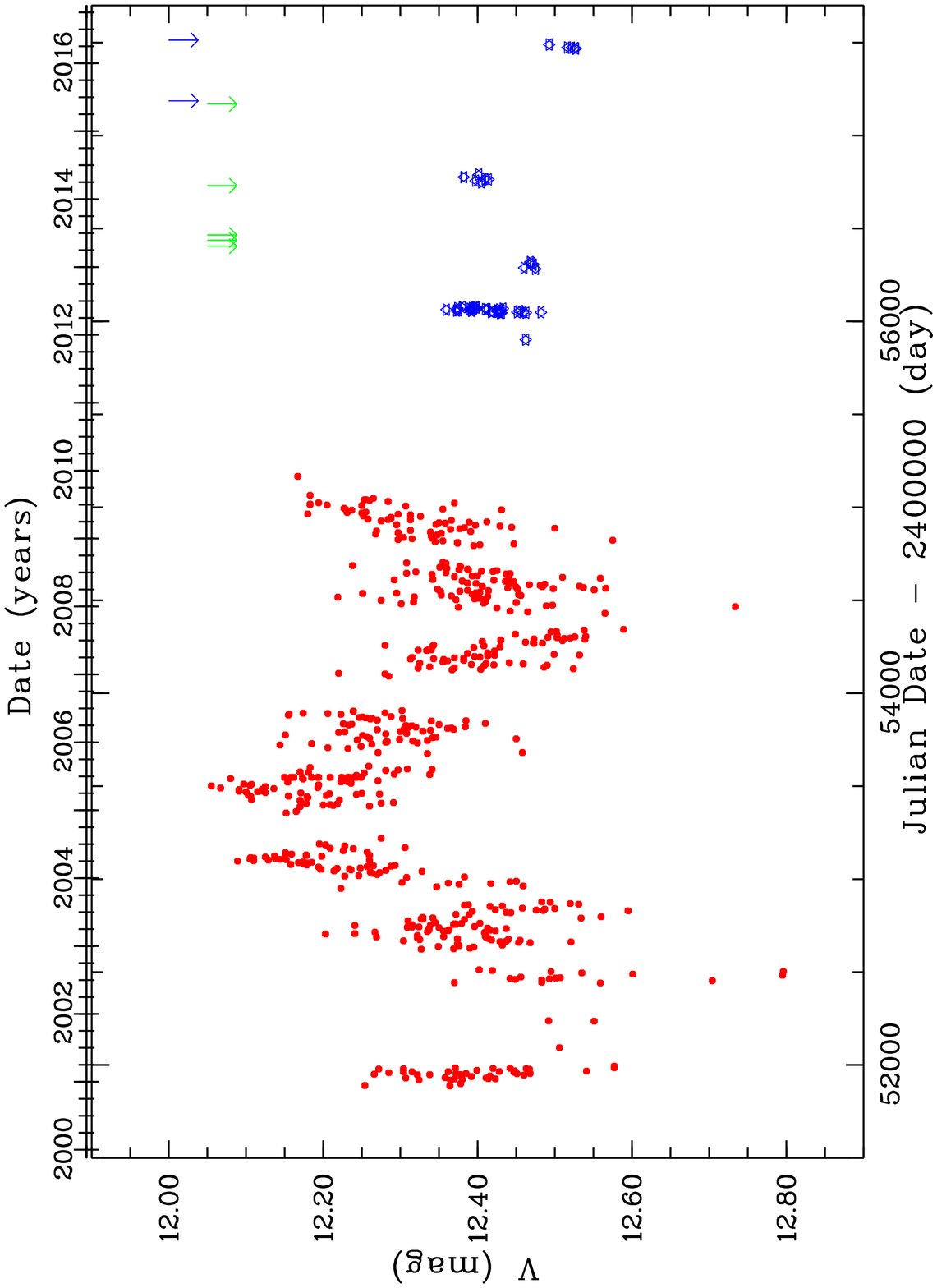}{fig:comparison}{A light
curve of Hen\,3$-$729 in the $V$-band in 1999--2016. The filled (red) data points are from the ASAS, while the open 
(blue) ones are based on our observations. The dates of the SALT RSS and HRS spectra are marked, respectively, by arrows
in the lower and upper rows.}

\articlefigure[angle=-90,width=0.9\textwidth,clip=]{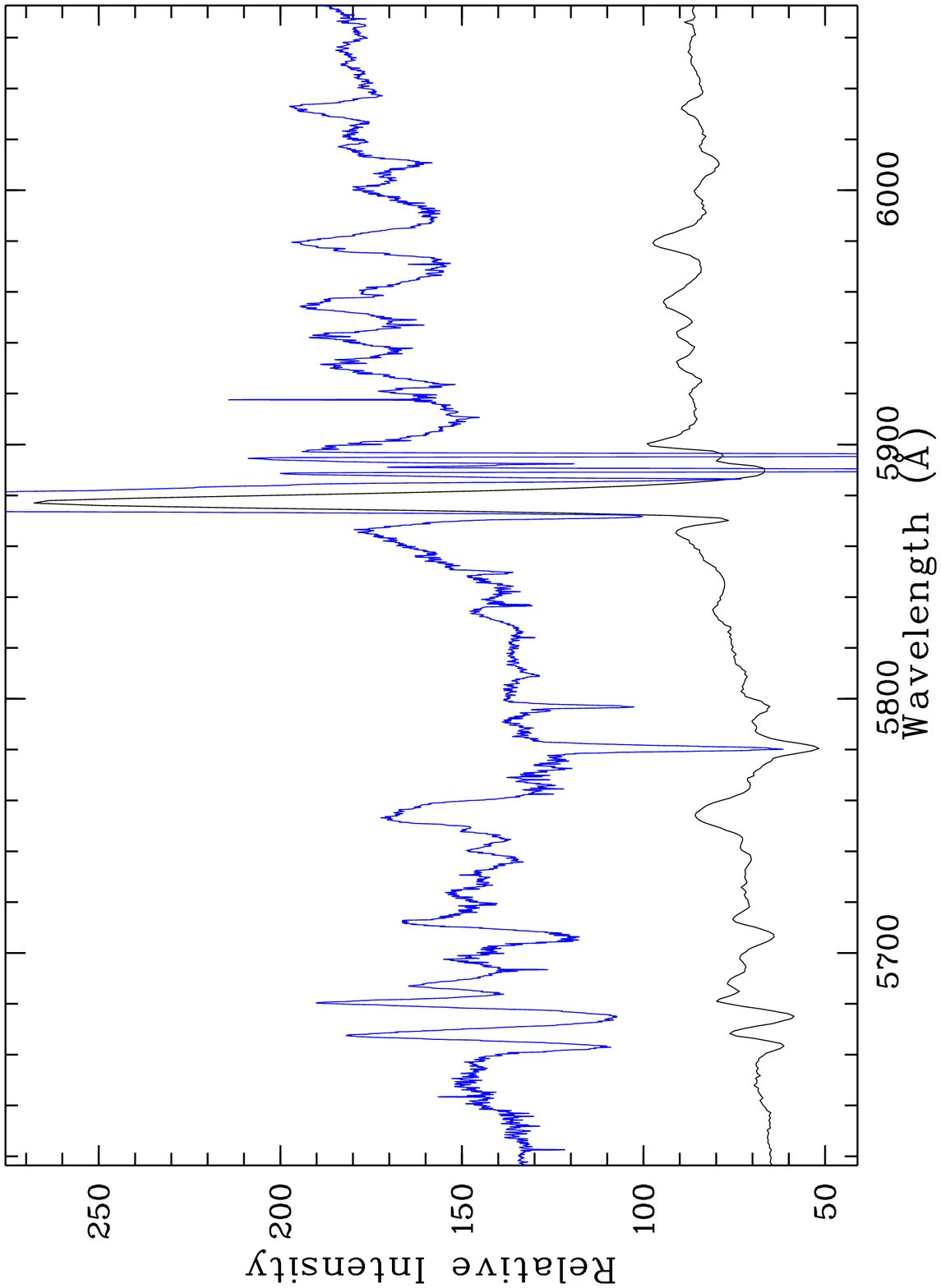}%
{fig:HRS_spec}{Comparison of a portion of the HRS spectrum (upper blue line) of the cLBV Hen\,3$-$729 with the corresponding part
of the portion of the RSS spectrum (bottom black line) of this star.}

\acknowledgements 
This work is based on observations obtained with the Southern African Large Telescope, programmes 
\mbox{2010-1-RSA\_OTH-001}, \mbox{2011-3-RSA\_OTH-002}, \mbox{2013-1-RSA\_OTH-014}, \mbox{2013-2-RSA\_OTH-003},
\mbox{2015-1-SCI-017} and  \mbox{2016-1-SCI-012}, and supported in part by the Russian Foundation
for Basic Research grant 16-02-00148. AYK acknowledges support from the National Research Foundation (NRF) of South 
Africa.



\end{document}